\documentclass{PoS}

\usepackage{amsfonts}
\usepackage{amsmath}
\usepackage{subfigure} 

\title{
  Determination of $B_K$ using improved staggered fermions
  (III) Finite volume effects 
}

\ShortTitle{
  Determination of $B_K$ using improved staggered fermions (III)
}

\author{\speaker{Boram Yoon}, Taegil Bae, Hyung-Jin Kim, Jangho Kim,
  Jongjeong Kim, Kwangwoo Kim, Weonjong Lee\\
  Frontier Physics Research Division and Center for Theoretical Physics \\
  Department of Physics and Astronomy, 
  Seoul National University, Seoul, 151-747, South Korea \\
  E-mail: \email{wlee@snu.ac.kr}}

\author{Chulwoo Jung \\
  Physics Department, Brookhaven National Laboratory,
  Upton, NY11973, USA \\
  E-mail: \email{chulwoo@bnl.gov}}

\author{Stephen R. Sharpe\\
  Physics Department, University of Washington, Seattle, WA 98195-1560 \\
  E-mail: \email{sharpe@phys.washington.edu}}

\author{Jisoo Yeo \\
  Posung High School, Bangi-dong, Songpa-gu, Seoul, 138-050, South Korea}

\abstract{ We study the finite-volume effects in our
calculation of $B_K$ using HYP-smeared improved staggered valence fermions.
We calculate the predictions of both
SU(3) and SU(2) staggered chiral perturbation theory at one-loop order.
We compare these to the results of a direct calculation,
using MILC coarse lattices with two different volumes: $20^3$ and
$28^3$.
From the direct calculation, we find that the finite volume effect is
$\approx 2\%$ for the SU(3) analysis and $\approx 0.9\%$ 
for the SU(2) analysis.
We also show how the statistical error depends on the
number of measurements made per configuration, and make a first
study of autocorrelations.}

\FullConference{
  The XXVII International Symposium on Lattice Field Theory - LAT2009\\
  July 26-31 2009
  Peking University, Beijing, China
}

\begin{document}

\section{Introduction} 
This paper is the third in a series of four reports on our calculation
of $B_K$ using HYP-smeared staggered fermions. 
In the previous two reports we presented the results of our analysis
using, respectively, SU(3) and SU(2) staggered chiral perturbation
theory (SChPT)~\cite{ref:wlee:2009-1,ref:wlee:2009-2}.
Here we explain how we estimate the finite-volume error, and
discuss issues concerning the statistical error.
The calculations use the MILC ensembles (C3) and (C3-2), the
parameters of which are given in Table~\ref{tab:fv}.

\begin{table}[h!]
\begin{center}
\begin{tabular}{ r | l }
\hline \hline
parameter           & value \\
\hline
$\beta$             & 6.76 ($N_f = 2+1$ unquenched QCD) \\
$1/a$               & 1.588(19) GeV \\
geometry            & $20^3 \times 64$ (C3) and $28^3 \times 64$ (C3-2) \\
configurations& 671 ($20^3$)  and  274 ($28^3$) \\
measurements per conf. & 9 ($20^3$)  and  8 ($28^3$) \\
sea quark (asqtad) masses      & $a m_l = 0.01$, $a m_s = 0.05$ \\
valence quark (HYP-smeared) masses& 0.005, 0.01, 0.015, \ldots, 0.04, 0.045, 0.05 \\
\hline \hline
\end{tabular}
\end{center}
\caption{Ensembles used to study finite volume effects.
  \label{tab:fv}}
\end{table}

\section{Finite Volume Correction: Theory}
Finite volume (FV) effects are predicted by ChPT, and are caused by
pseudo-Goldstone bosons (PGBs) in loops propagating from their starting
point to a periodic image. Experience indicates that the one-loop calculation
is a good guide to the sign and rough magnitude of the FV effect
(when compared to numerical results or higher order calculations),
although it tends to underestimate the size by factors of two or so.
We have calculated the one-loop FV effects using SChPT
for both SU(2) and SU(3) formulae.
These enter in the standard way
through the chiral logarithmic functions:
\begin{eqnarray}
\ell(X) &=& X \Big[ \log(X/\mu^2_\text{DR}) + \delta_1^\text{FV}(X,L) \Big]
\\
\tilde\ell(X) &=& - \frac{d \ell(X)}{dX} =
- \log(X/\mu^2_\text{DR}) - 1 + \delta_3^\text{FV}(X,L)
\end{eqnarray}
where $X=M^2$ is a PGB mass-squared (in physical units),
and the finite-volume corrections are given in terms of modified
Bessel functions by
\begin{eqnarray}
\delta_1^\text{FV} (X,L) &=& \frac{4}{z} \sum_{\vec{n} \ne 0} 
\frac{K_1 (|\vec{n}| z)}{|\vec{n}|}
\\
\delta_3^\text{FV}(X,L) &=& 2 \sum_{\vec{n} \ne 0}
K_0 (|\vec{n}| z)\,.
\end{eqnarray}
Here $z = ML$, with $L$ the spatial box size in physical units,
and $\vec{n}$ is a vector labeling the spatial image positions,
taking values $(1,0,0)$, $(0,1,0)$ etc..\footnote{%
Strictly speaking, we should also include images in the time direction as well,
but, since $L_t\gg L$ for the MILC lattices,
these contributions turn out to be negligible compared to the errors in $B_K$.}

When evaluating the image sums we 
find that keeping up to $\vec n^2=12$ in $\delta_1^{\rm FV}$
and $\vec n^2=16$ in $\delta_3^{\rm FV}$ gives sufficient accuracy.
As a side note, for our range of $z$, $2.5 \le z \le 10 $, it is
sometimes useful to have an {\em approximate} form to speed up the
calculation. We find that
\begin{eqnarray}
\delta^{\rm FV}_1(z) &\approx& e^{-z} z^{-3/2}
(26.0095 + 37.8217/z + 112.694/z^2 - 59.9524/z^3 )\,,
\\
\delta^{\rm FV}_3(z) &\approx& e^{-z} z^{-1/2}
(12.8717 + 10.7318/z + 88.9023/z^2 - 20.8847/z^3 )\,.
\end{eqnarray}
gives an accuracy of better than 1 part in $10^3$ over this range.
Note that these are {\em not} asymptotic expansions---these forms fail
for both small and large $z$.

In our fitting to SChPT forms (as described in the
companion reports~\cite{ref:wlee:2009-1,ref:wlee:2009-2}) we have, so far,
used the infinite-volume expressions. Thus it is important
to check that FV corrections are small.
Using the coefficients from the fit functions, we can calculate the
one-loop expectation for the FV corrections for each fit.
For the SU(3) fits, the largest contribution to the FV corrections
comes from terms induced by discretization and truncation errors.
These are not well determined, with the result that
different fits differ even in the sign of the predicted FV effect.
Thus we can at most use these calculations to estimate the magnitude
of the FV effect.

\begin{figure}[tbhp]
\centering
    \includegraphics[width=0.6\textwidth]{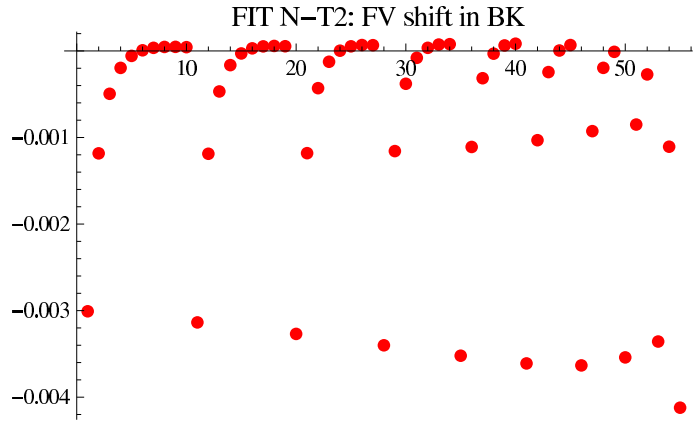}
\caption{Finite volume shift in $B_K$ as predicted by SU(3) SChPT 
for the $20^3\times 64$ coarse MILC lattice, 
using the ``N-T2'' fit~\protect\cite{ref:future}.
The horizontal axis orders the 55 valence ``kaons'' into groups
in which $a m_y$ is fixed and $am_x$ is increased from $0.005$ up
to $a m_y$. The first group (10 leftmost points) have $am_y=0.05
\approx a m_s^{\rm phys}$, the next group (9 points) have $a m_y=0.045$,
etc.. The rightmost point has $am_x=am_y=0.005$. The extrapolation to
the physical kaon mass is dominated by the leftmost group.
Note that the lowest ``row'' of points all have $am_x=0.005$.
}
\label{fig:su3chptfv}
\end{figure}
As an example, we show in Fig.~\ref{fig:su3chptfv}
the FV correction for the ensemble (C3) using
the fit-type which gives rise to the largest FV correction.
We see that the effect depends dominantly on the light-quark
valence mass $m_x$, and much more weakly on the strange valence mass
$m_y$. It grows rapidly as $m_x$ decreases, which is expected
given the exponential dependence on $z$.
The size of the FV effect is, however, small compared to $B_K\approx 0.5$---the
relative contribution does not reach 1\%.
We also note that on the larger, $28^3$, lattices, the FV correction
is smaller by almost an order of magnitude, and thus negligible compared
to other errors.

\begin{figure}[bthp]
\centering
    \includegraphics[width=0.49\textwidth]{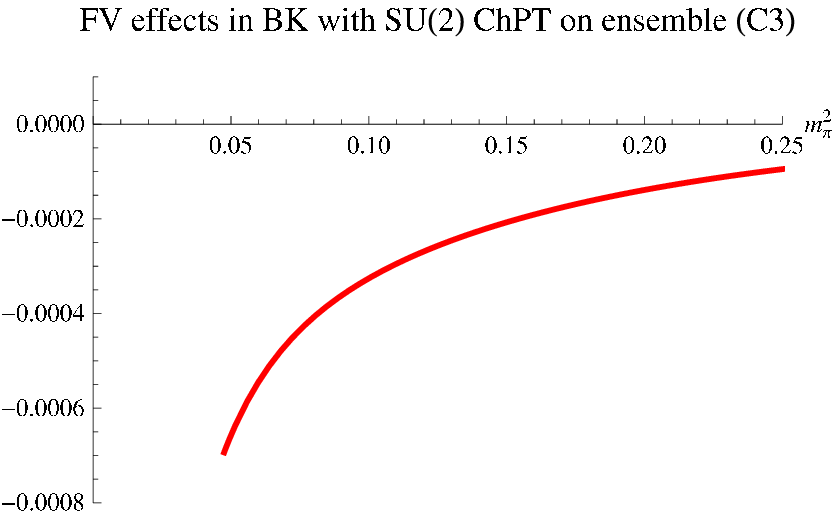}
    \includegraphics[width=0.49\textwidth]{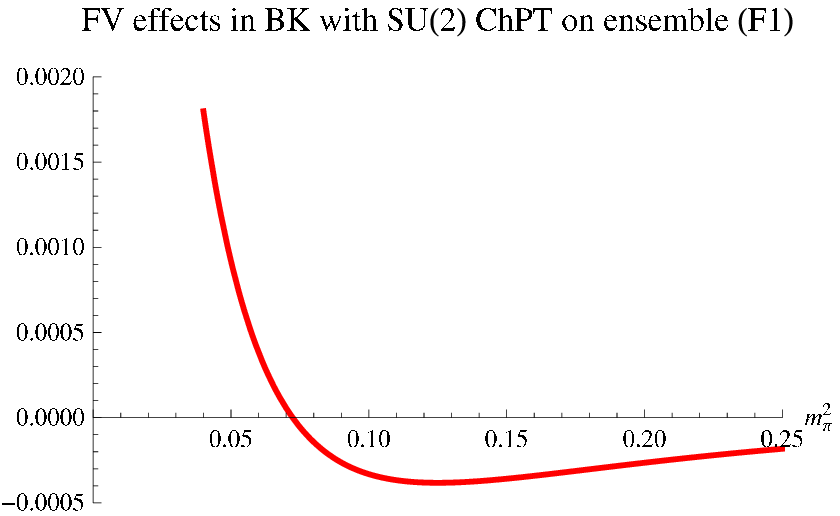}
\caption{Finite volume correction to $B_K$ from SU(2) SChPT as
a function of the squared mass of the pion composed of
light valence quarks, $M_\pi^2=X$ in GeV${}^2$.
Left panel: coarse ensemble (C3); right panel: fine ensemble (F1).
The lower limit of the curves corresponds to the
minimum values of $M_\pi^2$ in the simulations.}
\label{fig:su2-th-fv}
\end{figure}

The situation with SU(2) SChPT is shown in Fig.~\ref{fig:su2-th-fv},
here with results on both coarse and fine lattices.
The NLO prediction is more reliable in this case, since the
coefficient of the chiral logarithms is well determined by the fits.
We see again that the FV effect grows rapidly as $m_x\propto M_\pi^2$
decreases. The size of the effect is, however, very small,
significantly smaller than predicted by the SU(3) fit.
One peculiar feature is that the sign changes
(and magnitude increases) moving from coarse to fine lattices
(which are otherwise similar).
This is a reflection of the fact that, as one approaches the continuum limit,
taste splittings reduce, more pions become lighter, and FV effects increase.
In this specific case there are various canceling contributions and it
turns out also that the sign flips.

We draw two main conclusions from these results. First,
the FV correction to $B_K$ is small, comparable to or smaller than the 
statistical error for our most physical ``kaon''
(an error which is $\approx 0.6\%$ on the coarse ensemble).
Thus we are unlikely to make an error of larger than $\approx 1\%$
by leaving FV corrections out of the fits.
Second, we do not know in detail the expected form, or even the sign,
of the FV correction.
Note that the true FV correction for given valence masses has a definite value,
one that SU(2) and SU(3) ChPT should agree upon. The disagreement between the
two estimates is thus an indication that the NLO predictions are 
quantitatively unreliable.

This topic deserves further study. In particular we intend to
include one-loop FV corrections in future fits. In the meantime
we turn to a more direct method for estimated the FV error.

\section{Finite Volume Correction: Direct Measurement}

In order to study the finite volume effect numerically, we compare
results on the two coarse ensembles whose parameters are listed
in Table~\ref{tab:fv}. The lightest valence pion on the smaller lattice
has $m_\pi L \approx 2.68$. This is below the rule-of-thumb minimum ``safe''
value of 3-4, but we expect to have more leeway
when calculating kaon properties (such as $B_K$),
since the pion only enters through loops.
Indeed, the estimates of the previous section suggest that FV effects
remain small.
We also note that the same estimates indicate that FV effects are
almost negligible on the larger, $28^3$, lattices.

\begin{figure}[tbhp]
\centering
    \includegraphics[width=0.49\textwidth]{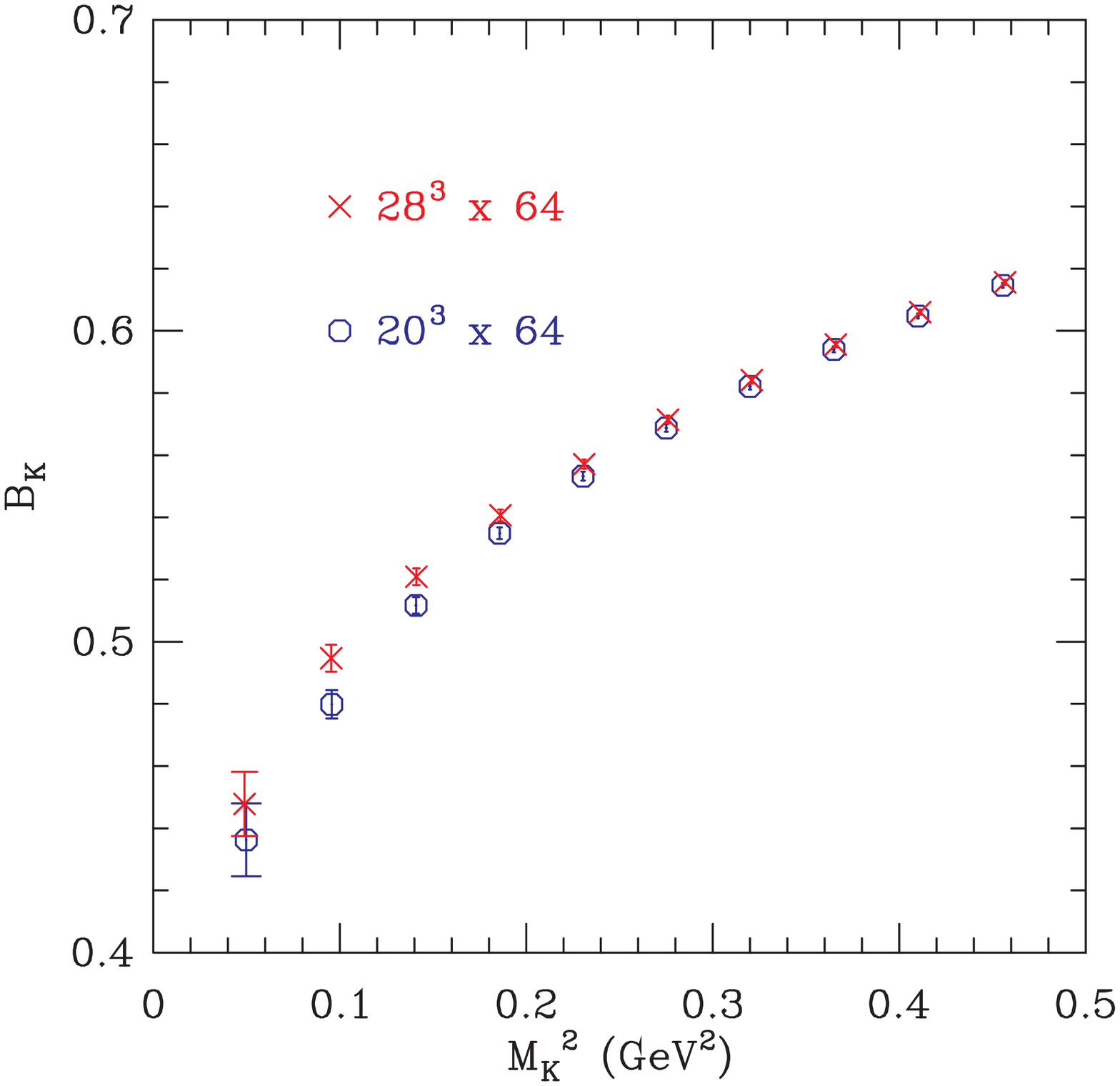}
    \includegraphics[width=0.49\textwidth]{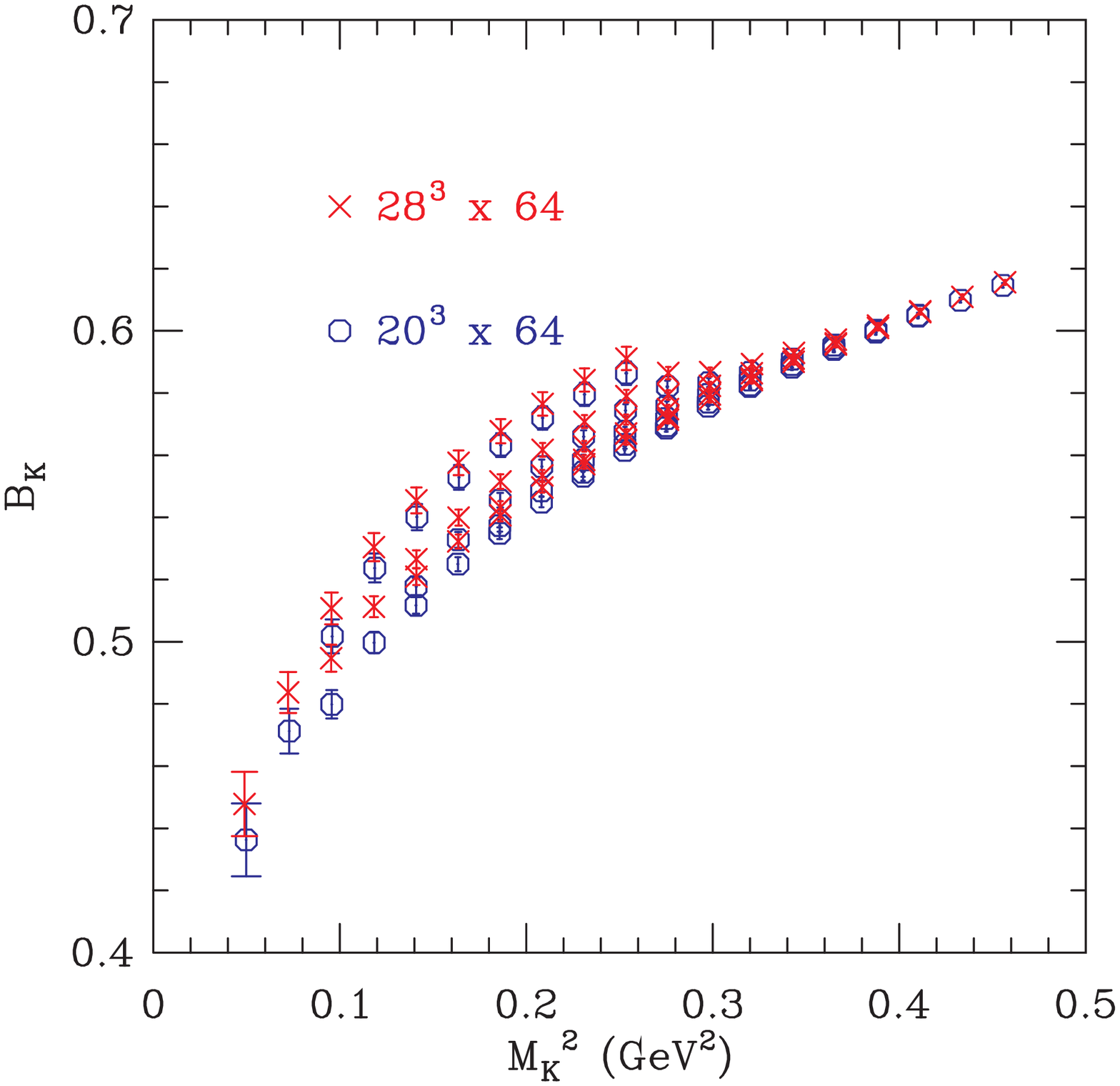}
\caption{ One-loop matched $B_K$ vs. $m_K^2$ for the 10
  degenerate combinations (left) and
  for all 55 degenerate and non-degenerate combinations (right).}
\label{fig:su3-fv}
\end{figure}

\begin{figure}[tbhp]
\centering
\includegraphics[width=0.49\textwidth]
                {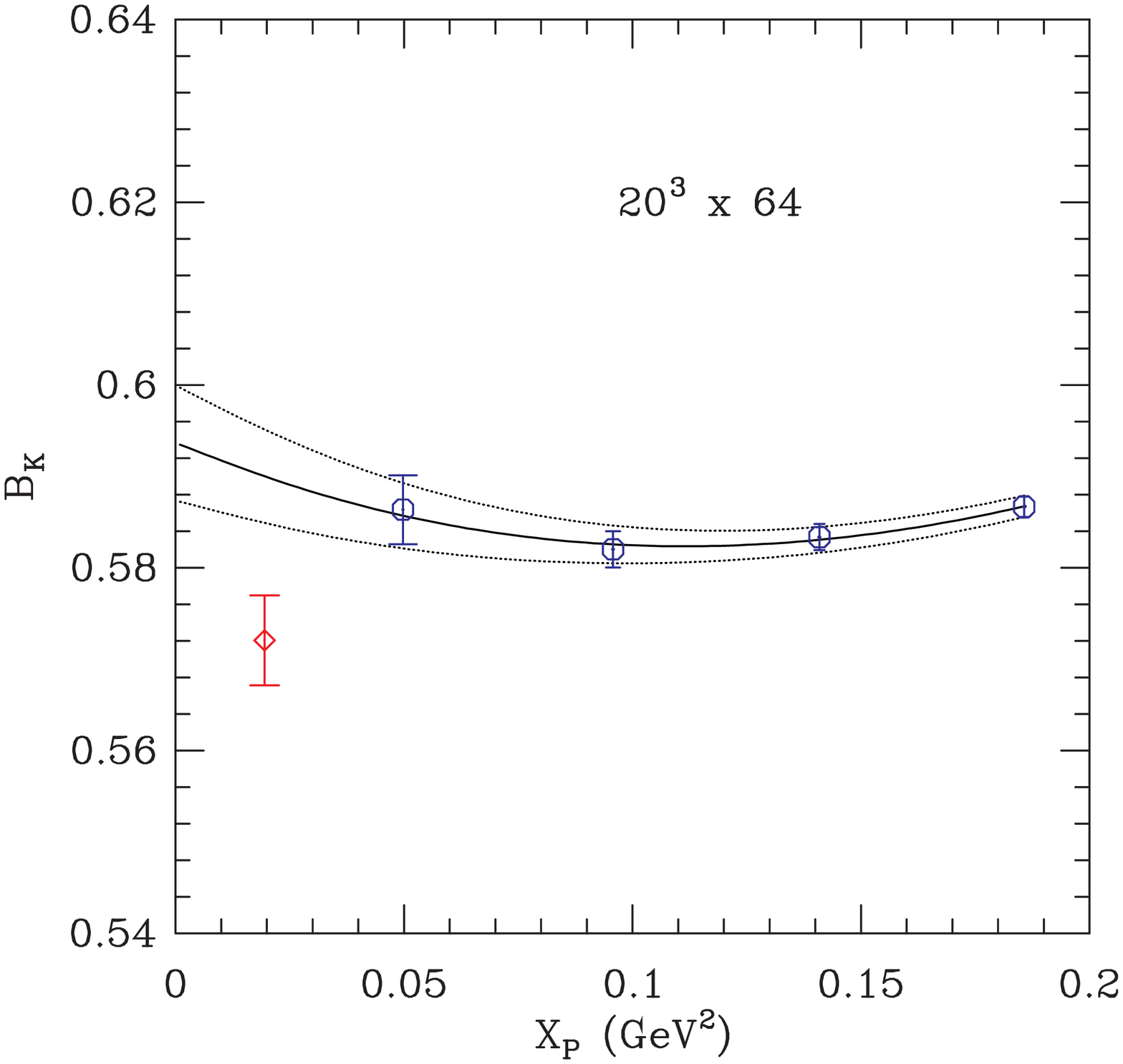}
\includegraphics[width=0.49\textwidth]
                {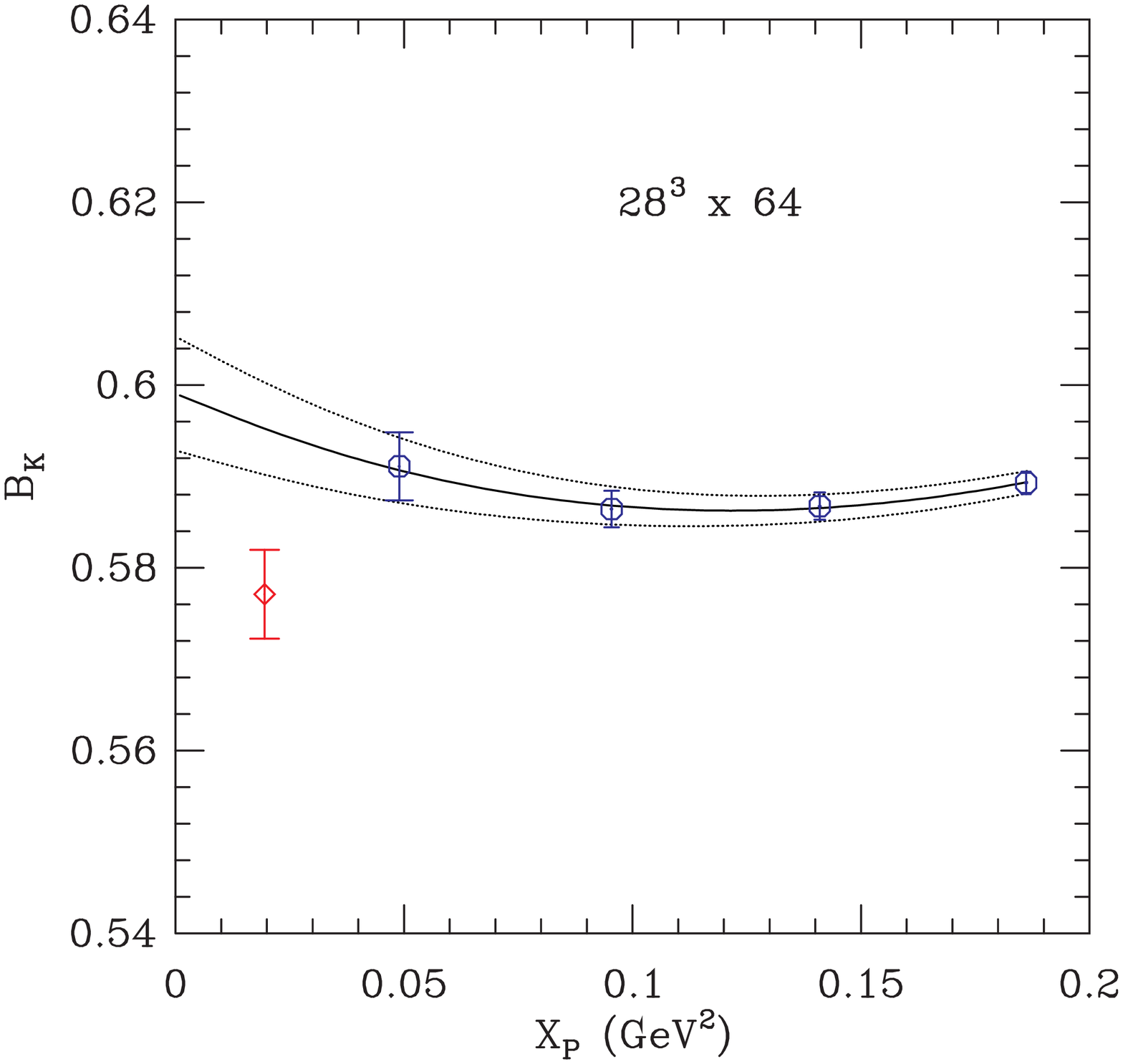}
\caption{One-loop matched $B_K$ vs. $X_P$ for the SU(2) analysis on the $20^3$ volume
  (left) and on the $28^3$ volume (right). Here $X_P$ is the mass-squared of
the Goldstone taste ``pion'' composed of light valence quarks of mass
$a m_x$. See Refs.~\cite{ref:wlee:2009-1,ref:wlee:2009-2} for more details.}
\label{fig:su2-fv}
\end{figure}

We have adjusted the number  of
measurements per configuration so that the statistical
weights of the two ensembles are very similar:
\begin{equation}
\textrm{ratio of stat. weights} 
= \frac{20^3 \times 671 \times 9}{28^3 \times 274 \times 8} = 1.004\,.
\end{equation}
In Fig.~\ref{fig:su3-fv}, we show the results for $B_K$ on the two
lattices, and find that the errors are indeed comparable.
The results on the larger lattice lie systematically higher,
although for each point the difference is only about 1 $\sigma$.
%
%
%
%
We fit both data-sets using our preferred Bayesian ``N-BT7'' fit
for the SU(3) analysis, and the ``4X3Y-NNLO'' fit for the
SU(2) analysis.\footnote{%
The details of these fits are discussed in 
Refs.~\cite{ref:wlee:2009-1,ref:wlee:2009-2}.}
The SU(2) ``X-fits'' are shown in Fig.~\ref{fig:su2-fv}.
The results of the fits are summarized in Table \ref{tab:su3-su2-fv}.
The difference between $B_K$ on the two
volumes is $\approx 2\%$ for the SU(3) analysis, 
which is about twice the statistical error, and thus probably significant.
For the SU(2) analysis the finite volume effect is smaller, $\approx 0.9\%$,
and, being comparable to the statistical error, less significant.
We interpret the difference between SU(3) and SU(2) analyses as being
due to greater simplicity of the SU(2) fit form.
The size of these effects are not unexpected given the theoretical
results of the previous section.
We use these differences as estimates of the FV systematic error
in the respective fits, as quoted in Refs.~\cite{ref:wlee:2009-1,ref:wlee:2009-2}.

\begin{table}[h]
\begin{center}
\begin{tabular}{ c | c  c}
\hline \hline
volume  &  SU(3)  &  SU(2)  \\
\hline
$20^3$  &  0.5599(57) & 0.5645(49) \\
$28^3$  &  0.5730(57) & 0.5694(48) \\
\hline \hline
\end{tabular}
\end{center}
\caption{Results for $B_K$ using SU(3) and SU(2) staggered ChPT.
  The values quoted are in the NDR scheme at a scale of $\mu=2\;$GeV.
  \label{tab:su3-su2-fv}}
\end{table}

\section{Increasing the Statistics and Autocorrelations}
\begin{figure}[htbp]
  \centering
  \includegraphics[width=0.5\textwidth]{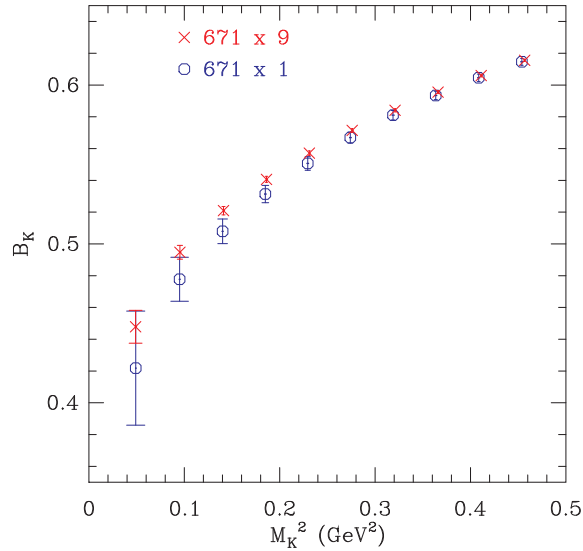}
\caption{ $B_K$ vs. $m_K^2$ on ensemble (C3) for one measurement/conf. and
  for 9 measurements/conf.}
\label{fig:high-stat-1}
\end{figure}

It is well know that one can obtain more information from each
gauge configuration by using multiple time-positions for the sources.
An important issue, however, is the extent to which these measurements
are independent.
We have tested this by using 9 randomly chosen ``starting times''
for the wall sources on each configuration in the coarse ensemble (C3),
each with a different random seed for the noise in the source.
We average the raw data from these 9 sources,
and then compute the error using single elimination jackknife.
We compare the results so obtained (labeled ``x9'') to those
obtained using a single measurement/configuration (``x1'')
in Fig.~\ref{fig:high-stat-1}, displaying only degenerate
combinations for clarity.
We observe that the errors are reduced by a factor of 3, indicating
that the different measurements on each configuration are almost
independent.

\begin{figure}[htbp]
  \centering
  \includegraphics[width=0.49\textwidth]{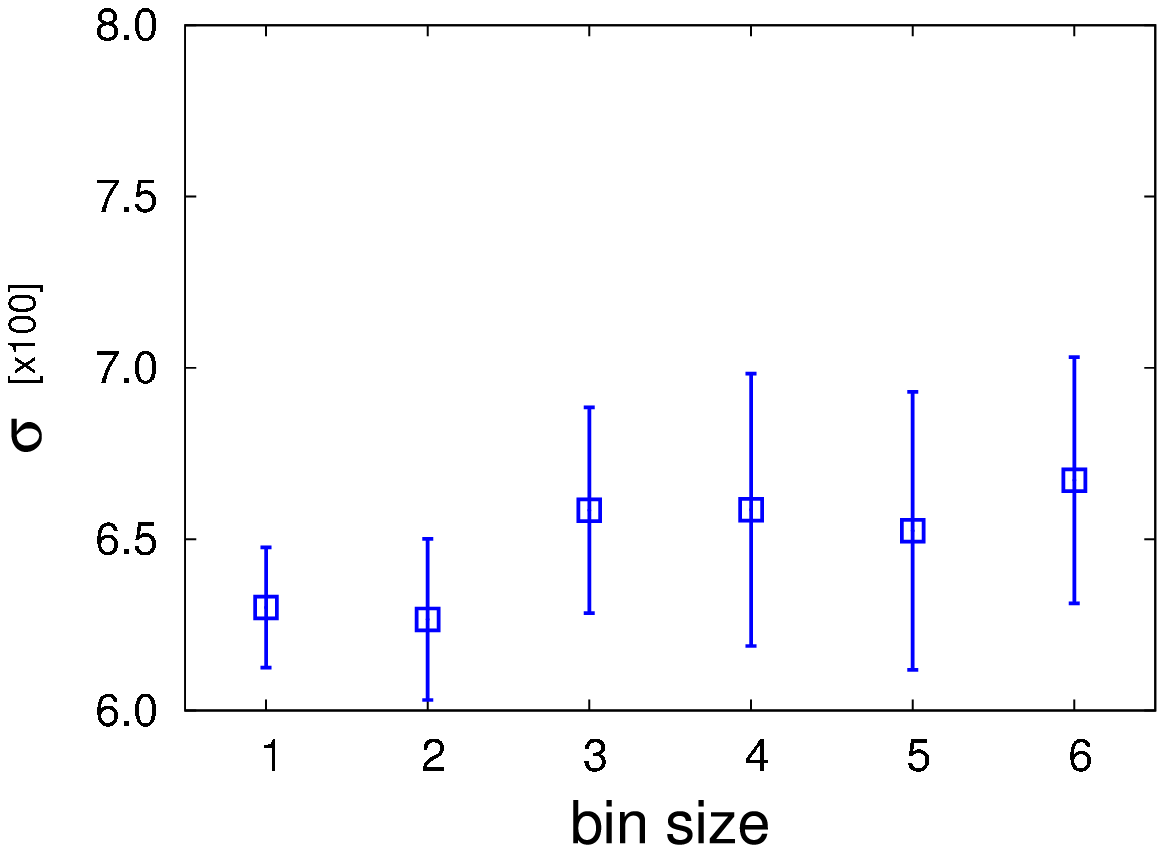}
  \includegraphics[width=0.49\textwidth]{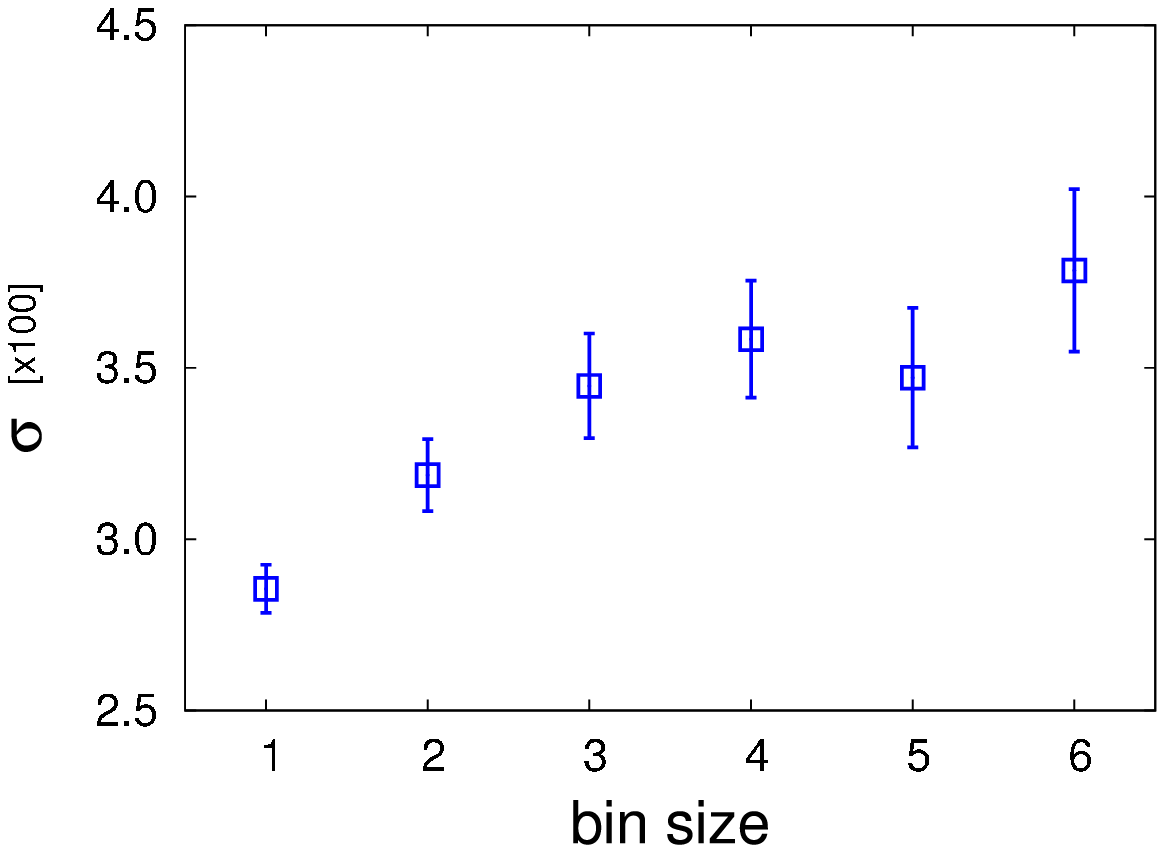}
\caption{ Error in $C(t=10)$ [see text] vs. bin size 
for the x1 data (left) and the x9 data (right).}
\label{fig:autocorr}
\end{figure}

We have also studied autocorrelations between configurations by
looking at the dependence of the statistical errors on the bin size
(i.e. the number of configurations we bin together)
In Fig.~\ref{fig:autocorr}, we display the result for
the  error on the 2-point correlator
\begin{eqnarray}
C(t) = \langle A_4(t) P(0) \rangle
\end{eqnarray}
(where both operators have the Goldstone taste $\xi_5$)
at $t=10$.
For the x1 data, the error in the error bar is too large to
determine whether there is an autocorrelation.\footnote{%
Details of how we estimate the error in the error bar 
will be given in Ref.~\cite{ref:future}.}
For the x9 data, however, we do observe about a 20\% increase
between a bin size of 1 and 3-6.
Fortunately this is a small effect.
Nevertheless, it is clear that having higher statistics
allows us to better understand the autocorrelations,
and we intend to increase the statistics on other ensembles as well.

\section{Acknowledgments}
C.~Jung is supported by the US DOE under contract DE-AC02-98CH10886.
The research of W.~Lee is supported by the Creative Research
Initiatives Program (3348-20090015) of the NRF grant funded by the
Korean government (MEST). 
The work of S.~Sharpe is supported in part by the US DOE grant
no.~DE-FG02-96ER40956. Computations were carried out
in part on facilities of the USQCD Collaboration,
which are funded by the Office of Science of the
U.S. Department of Energy.

\end{document}